\documentclass[conference]{IEEEtran}

\usepackage{cite}

\ifCLASSINFOpdf
\else
\fi

\usepackage{amsmath}
\usepackage{array}

\usepackage{wasysym}
\usepackage{enumitem}
\usepackage{todonotes}
\usepackage{algorithm} %
\usepackage{algpseudocode}
\usepackage{threeparttable}
\usepackage{booktabs}
\usepackage{hyperref}
\usepackage[capitalise,noabbrev]{cleveref}
\usepackage{multirow}
\usepackage{makecell}
\usepackage{amssymb}

\newcommand{\etal}[1]{#1 \emph{et al.}}
\newcommand{\schemename}{\textsc{Silenzio}}

\newcommand{\ZZ}{\mathbb{Z}}
\newcommand{\LineComment}[1]{\Statex\(\triangleright\) \textit{#1}}
\newcommand{\MyComment}[1]{\Comment{\textit{#1}}}

\begin{document}
\title{\schemename: Secure Non-Interactive Outsourced MLP Training}

\author{\IEEEauthorblockN{Jonas Sander}
	\IEEEauthorblockA{University of Luebeck\\
		j.sander@uni-luebeck.de}
	\and
	\IEEEauthorblockN{Thomas Eisenbarth}
	\IEEEauthorblockA{University of Luebeck\\
		thomas.eisenbarth@uni-luebeck.de}}

\maketitle

\begin{abstract}
Outsourcing ML training to cloud-service-providers presents a compelling opportunity for resource constrained clients, while it simultaneously bears inherent privacy risks.  We introduce \schemename, the first fully non-interactive outsourcing scheme for the training of MLPs that achieves 128\,bit security using FHE (precisely TFHE). Unlike traditional MPC-based protocols that necessitate interactive communication between the client and server(s) or non-collusion assumptions among multiple servers, \schemename{} enables the ``fire-and-forget'' paradigm without such assumptions. In this approach, the client encrypts the training data once, and the server performs the training without any further interaction.

\schemename{} operates entirely over low-bitwidth integer to mitigate the computational overhead inherent to FHE. Our approach features a novel low-bitwidth matrix multiplication gadget that leverages input-dependent residue number systems, ensuring that no intermediate value overflows 8\,bit. Starting from an RNS-to-MRNS conversion process, we propose an efficient block-scaling mechanism, which approximately shifts encrypted tensor values to their user-specified most significant bits. To instantiate the backpropagation of the error, \schemename{} introduces a low-bitwidth gradient computation for the cross-entropy loss.

We evaluate \schemename{} on standard MLP training tasks regarding runtime as well as model performance and achieve similar classification accuracy as MLPs trained using PyTorch with 32\,bit floating-point computations. Our open-source implementation of \schemename{} represents a significant advancement in privacy-preserving ML, providing a new baseline for secure and non-interactive outsourced MLP training.
\end{abstract}

\IEEEpeerreviewmaketitle

\section{Introduction}
Machine learning (ML) and particularly the rise of powerful artificial neural networks (NNs), has transformed numerous industries through significant breakthroughs in pattern recognition, decision-making, and predictive analytics. Among these, Multi-Layer Perceptrons (MLPs) constitute a foundational and versatile class of NNs, widely employed in critical areas ranging from medical diagnostics to financial analytics. However, the ever-increasing complexity of ML models and size of datasets necessitate substantial computational resources that are often unavailable to resource-constrained clients, such as individual users (e.g., IoT or medical devices) or small-scale enterprises. Consequently, outsourcing ML training tasks to cloud-service-providers emerges as a practical solution.

Despite the clear advantages of outsourced NN training, serious concerns regarding privacy and security arise, especially when the underlying datasets contain sensitive personal or economically valuable information. Traditional solutions have predominantly utilized Multi-Party Computation (MPC)-based protocols, e.g., \cite{DBLP:conf/uss/Dalskov0K21, DBLP:conf/uss/KotiPPS21, DBLP:conf/uss/WatsonWP22, DBLP:conf/uss/RatheeB00S23, DBLP:conf/uss/RenF00ZHH024, DBLP:conf/uss/LycklamaVKKH24, DBLP:conf/sp/TanKTW21, DBLP:journals/popets/WaghTBKMR21, DBLP:journals/popets/AttrapadungHIKM22, DBLP:journals/popets/BaccariniBY23, DBLP:journals/corr/abs-2403-17296, DBLP:conf/ndss/KotiPRS22}, relying heavily on interactive communication or strict non-collusion assumptions among multiple servers. Such requirements introduce significant practical limitations, hindering their widespread adoption in scenarios demanding secure, seamless and scalable solutions.

Existing non-interactive schemes for NN training not requiring non-collusion assumptions \cite{DBLP:conf/cvpr/NandakumarRPH19, DBLP:conf/nips/LouFF020} solely rely on Fully Homomorphic Encryption (FHE), but only guarantee around 80\,bits of security in their evaluation to significantly reduce their computational overhead and are therefore considered insecure by today's standards. Furthermore, there is no open-source implementation of such schemes, massively hindering further advancements in the development of practical and secure solutions for non-interactive outsourcing of ML training.

To address the need for  secure and  fully non-interactive outsourced training of MLPs, we introduce \schemename. \schemename{} solely relies on FHE — particularly Fast Fully Homomorphic Encryption over the Torus (TFHE) \cite{DBLP:journals/joc/ChillottiGGI20} — and provides a robust security guarantee of 128\,bits.
To meet performance requirements and  minimize the computational overhead, \schemename{} draws from recent advancements in hardware-accelerated ML \cite{DBLP:journals/tpds/WangRLS22} and shows how similar approaches can accelerate TFHE-protected ML training. \schemename{} leverages exclusively low-bitwidth integer arithmetic, never exceeding 8\,bits, thus significantly mitigating the computational overhead characteristic of FHE-protected computations. As part of \schemename, we propose three new building blocks to enable effective training without exceeding the 8\,bit limit for all FHE computations. First, we introduce a low-bitwidth matrix multiplication gadget leveraging input-dependent residue number systems (RNS) and a low-bitwidth modular summation routine.
Based on a new vectorized and FHE-protected implementation for RNS to mixed-radix number system (MRNS) conversions, we propose the second building block: a novel block-scaling gadget that approximately shifts the values of an encrypted input tensor given in RNS representation to its user specified $\Gamma$ most significant bits. As the third and final building block, we propose a simple approximated gradient computation for the cross-entropy loss compatible with TFHE.

Our implementation of \schemename{} is based on Zama's state-of-the-art Concrete library \cite{Concrete}. We comprehensively evaluate \schemename{} using standard benchmark datasets and various MLP configurations, establishing a new performance baseline for privacy-preserving outsourced MLP training. Furthermore, \schemename{} will be released as open-source, aiming to encourage further research regarding the non-interactive outsourcing of NN training.

\subsection{Contributions}
We introduce \schemename, the first fully non-interactive training scheme for MLPs that provides full-strength security while achieving  unprecedented performance due to the following contributions:
\begin{itemize}[noitemsep,leftmargin=*]
\item  \schemename\ achieves 128\,bit of security while only requiring low-bitwidth integer computations and never exceeding 8\,bit for FHE-computations. 
\item A new low-bitwidth matrix-multiplication gadget for up to 8\,bit input-matrices that leverages input-dependent residue number systems, a low-bitwidth summation routine, and optionally a Karatsuba-inspired multiplication engine to keep all FHE-processed values, including the output, within 8\,bit value ranges.
\item A new FHE-protected (block-)scaling gadget, $\text{Shift2MSBs}^{\pm}$, that approximately shifts the values of an input-tensor represented in a residue number system to its user-specified $\Gamma$ most significant bits.
\item As part of $\text{Shift2MSBs}^{\pm}$, we provide a fast FHE-protected implementation for number conversions from residue number systems to associated mixed-radix number systems.
\item A low-bitwidth and TFHE-friendly gradient computation for the cross-entropy loss.
\item \schemename's end-to-end training approach is implemented using the state-of-the-art Concrete library, providing good extensibility and a low-barrier starting point for further research. To the best of our knowledge, we provide the first open-source implementation of a non-interactive (without non-collusion assumption) cryptographic outsourcing scheme for MLP training. 
\end{itemize}
We will release the code upon acceptance.

\section{Preliminaries}
We note vectors with bold lower-case (e.g., $\mathbf{m}$) and matrices respective tensors with bold upper-case letters (e.g., $\mathbf{W}$). $\ZZ_q$ denotes the ring of integer modulo $q$ and $\lceil\cdot\rceil$ rounding upwards to the next integer.  We summarize used notations in \cref{tab:notation}.

\subsection{Non-Interactive Outsourcing}

Cryptographic neural network computations are broadly categorized into four distinct scenarios, each differing substantially in terms of interactivity, assumptions, and practical constraints (see also \cite{DBLP:conf/sp/NgC23}). To further motivate \schemename's scenario and point out the differences to other settings, we shortly introduce them in the following.

\textit{Oblivious Inference/Training} schemes typically involve a single server providing inference or training services without learning the client's input data~\cite{DBLP:conf/ccs/HussainJSK21, DBLP:conf/ccs/BallaK23, DBLP:conf/uss/LehmkuhlMSP21, DBLP:conf/uss/NgC21, DBLP:conf/uss/Chandran0OS22, DBLP:conf/uss/HuangLHD22, DBLP:conf/uss/LiuX024, DBLP:conf/sp/RatheeB00CR22, singh2024hyena, zhang2024individual, DBLP:journals/popets/GuptaKCG22, DBLP:journals/popets/VeldhuizenSVK24, DBLP:conf/ndss/ZhangXW21, DBLP:conf/ndss/Dong0L0TYCH23, GuptaCGKS25}. Although effective for protecting inputs, these schemes inherently rely on frequent client-server interactions beyond initial setup, leading to significant computational and communication overheads, often outweighing the efficiency benefits of cloud outsourcing. \textit{Private Inference/Training} addresses scenarios involving multiple data owners who collaboratively compute inference results or train models without mutually revealing sensitive input data~\cite{DBLP:conf/ccs/SongWWTLRWH22, DBLP:conf/uss/Patra0SY21, DBLP:conf/uss/WatsonWP22, DBLP:conf/uss/LiDHHZS23, DBLP:conf/uss/YuanYZ0G024, DBLP:conf/sp/TanKTW21, DBLP:conf/sp/TianZRCZ0022, jawalkar2023orca, DBLP:journals/popets/Wagh22, DBLP:journals/popets/BaccariniBY23, DBLP:journals/popets/DasCCGLS25, DBLP:conf/ndss/SavPTFBSH21, DBLP:journals/corr/abs-2403-11166, biswas2025low, harth2025pigeon}. However, these approaches generally require substantial inter-party communication, making them unsuitable for clients seeking minimal online involvement.

To alleviate these communication-intensive requirements, \textit{Semi-Non-Interactive Outsourced Inference/Training} schemes like e.g., \cite{DBLP:conf/uss/Dalskov0K21, DBLP:conf/uss/KotiPPS21, DBLP:conf/uss/WatsonWP22, DBLP:conf/uss/RatheeB00S23, DBLP:conf/uss/RenF00ZHH024, DBLP:conf/uss/LycklamaVKKH24, DBLP:conf/sp/TanKTW21, DBLP:journals/popets/WaghTBKMR21, DBLP:journals/popets/AttrapadungHIKM22, DBLP:journals/popets/BaccariniBY23, DBLP:journals/corr/abs-2403-17296, DBLP:conf/ndss/KotiPRS22, harth2025pigeon} adopt a more cloud-compatible "fire-and-forget" paradigm. In these protocols, the client initially secret-shares their input among multiple independent servers, after which they may go offline until computation completion. This approach significantly reduces online communication but introduces critical security assumptions: the client must fully trust that servers do not collude, which might be difficult to guarantee — particularly when servers belong to the same organization or jurisdiction. To eliminate reliance on potentially unrealistic non-collusion assumptions, \schemename{} leverages the \textit{Non-Interactive Outsourced Inference/Training} setting (for an overview, see \cref{tab:featuresetcomparison} and \cref{sec:related_work}). Here, the client encrypts their data using cryptographic primitives — such as FHE — and sends the ciphertext to a single server. After initial transmission, no further interaction between the client and server is required until the server completes its computation. This setting entirely removes the need for non-collusion assumptions, making it a straightforward ``drop-in replacement'' for conventional, unprotected cloud offerings. In some sense, this setting is also the counterpart to typical computations protected through trusted execution environments (TEEs). In comparison, \schemename{} does not rely on the additional hardware-based security assumptions bound to TEEs and are often shown to be vulnerable, especially through side-channel attacks, like e.g., \cite{DBLP:conf/sosp/BulckPS17, DBLP:conf/ccs/WilkeS024}.

\subsection{Multi-Layer Perceptrons}

Multi-Layer Perceptrons (MLPs) are a class of artificial neural network (NN) that consists of multiple layers of neurons, organized in a feedforward architecture. MLPs serve as universal function approximators and are widely used in various machine learning tasks, including classification and regression. In deep learning, MLPs serve as foundational building block for more complex architectures such as convolutional NNs and recurrent NNs.

An MLP is composed of an input layer, one or more hidden layers, and an output layer. Each layer consists of multiple neurons, which apply an affine transformation followed by a nonlinear activation function. Mathematically, for a given layer $l$, the output $\mathbf{h}^{(l)}$ is computed as: $\mathbf{h}^{(l)} = \sigma(\mathbf{W}^{(l)} \mathbf{h}^{(l-1)} + \mathbf{b}^{(l)})$
where $\mathbf{W}^{(l)}$ represents the weight matrix, $\mathbf{b}^{(l)}$ is the bias vector, and $\sigma(\cdot)$ denotes a nonlinear activation function such as the $\operatorname{ReLU}$ function $\operatorname{ReLU} = \max(0, x)$.

Training an MLP involves adjusting the weights and biases to minimize a loss function, typically through backpropagation and gradient-based optimization methods like stochastic gradient descent (SGD).

\subsection{Fully Homomorphic Encryption}
\label{sec:fhe}
Fully Homomorphic Encryption is a form of encryption that allows computations to be performed directly on encrypted data without requiring decryption. This property is particularly useful for privacy-preserving computations, enabling secure outsourcing of computations to untrusted environments such as cloud-service-providers. Most practical FHE schemes are based on the Learning-With-Errors (LWE) or Ring-Learning-With-Errors (RLWE) problems, which add an error term to the secret to protect its confidentiality. FHE schemes support arbitrary computations on ciphertexts by enabling both addition and multiplication operations. Performing such arithmetic operations on ciphertexts accumulates the added noise until the ciphertexts are not decryptable anymore. To allow an arbitrary number of operations on a ciphertext, FHE schemes provide a so-called bootstrapping operation that reduces the noise and allows for more arithmetic operations to be performed. An FHE scheme typically consists of the following suite of algorithms:
\begin{itemize}[noitemsep,topsep=0pt]
    \item \textbf{Key Generation} $\operatorname{KeyGen}(\lambda) \rightarrow (pk, sk)$: Outputs a public evaluation and secret en-/decryption key based on the given security parameter $\lambda$.
    \item \textbf{Encryption} $\operatorname{Enc}(p,sk) \rightarrow c$: Transforms a plaintext input into a ciphertext, using the secret encryption key.
    \item \textbf{Evaluation} $\operatorname{Eval}_f(c, pk) \rightarrow c'$: Given the public evaluation key allows a function $f$ to be performed on ciphertexts, generating encrypted results equivalent to those obtained by performing the same function on the plaintexts.
    \item \textbf{Decryption} $\operatorname{Dec}(sk, c) \rightarrow m$: Converts the computed ciphertext back into plaintext using the secret key.
\end{itemize}

\subsubsection{TFHE \& Concrete}
We implement \schemename{} using Zama's Concrete library, that provides a Python-based interface to construct arbitrary integer-based circuits, which then are compiled to efficient TFHE-protected programs. While the circuit itself is arbitrary, Concrete is limited to 8\,bit integer and using RNS representations under the hood up to 16\,bit integer computations. TFHE's bootstrapping is programmable, meaning it allows to perform arbitrary operations on the inputs during the bootstrapping process. Based on TFHE's programmable bootstrapping (PBS), Concrete provides next to the regular linear operations addition, subtraction, and multiplication, also non-linear operations modeled through table lookups. In Concrete, the security parameter $\lambda$ is fixed to 128\,bit.

\subsection{Number Systems}
To keep the operands of FHE operations in manageable value ranges, \schemename{} relies on the following number systems.

\subsubsection{Residue Number System (RNS)}
    The residue number system (RNS) is a non-weighted number system that represents integer using a set of relatively prime moduli. Given a set of moduli $\{m_1, m_2, \dots, m_k\}$, called the RNS base, an integer $x$ is uniquely represented by the tuple $(x_1, x_2, \dots, x_k)$, where: $x_i = x \bmod m_i, \text{ for } i \in [k]$. Given this base, the RNS has a cardinality of $\Pi_{i=1}^k m_i = M_k$, meaning it can represent $M_k$ distinct values uniquely.

Operations in an RNS can be performed independently on the individual residues without the need to handle intermediate carry values, making the RNS predestined for use with FHE schemes. More specifically, using an RNS allows breaking up large numbers into multiple smaller residues, accelerating FHE operations. RNS representations are widely used to speed up FHE-protected computations \cite{DBLP:conf/sacrypt/CheonHKKS18, DBLP:journals/iacr/BonehK25}.

\subsubsection{Mixed-Radix Number System (MRNS)}
The Mixed-Radix Number System (MRNS) extends traditional positional numbering by allowing each digit to have a varying base. A number $x$ in an MRNS with the radix-base $\{r_1, r_2, \dots, r_k\}$ is represented as $x = x_1+ \sum_{i=2}^{k} x_i \prod_{j=1}^{i-1} r_j$ where the digits $x_i$ satisfy $0 \leq x_i < r_i$. Similar to the RNS system, the cardinality of the MRNS is $R_k = \Pi_{i=1}^k r_i$.

\subsubsection{RNS-to-MRNS Conversion}
A value represented in RNS can be converted into an associated MRNS \cite{szabo1967residue}. An RNS and an MRNS are called associated if for the set of moduli $\{m_0, m_1, \dots, m_k\}$ that define the RNS and the set of radices $\{r_0, r_1, \dots, r_k\}$ that define the MRNS we have $\forall i \in [k]: m_i = r_i$. %
\Cref{lst:rns_to_mrs_conversions} shows the conversion process of a number $x$ given in RNS representation to a number $y=x$ in an associated MRNS representation, as described by Szabó and Tanaka \cite{szabo1967residue}. \Cref{tab:rns_to_mrs_example} shows an example.

\begin{algorithm}
\caption{RNS2MRNS}
\label{lst:rns_to_mrs_conversions}
\begin{algorithmic}
    \Require $(x_1,\ldots , x_k)$, $(m_1,\ldots,m_k)$ \MyComment{$x$ in RNS, RNS base}
    \Ensure $(y_1,\ldots , y_k)$, $(r_1,\ldots,r_k)$ \MyComment{$x$ in MRNS, MRNS base}
    \State $(y_1,\ldots,y_k)\gets (x_1,\ldots,x_k)$
    \For {$i \gets 2 \text{ to } k$}
        \State $(y_i,\ldots,y_k) \gets (y_i,\ldots,y_k) - y_{i-1}$
        \State $(t_i,\ldots,t_k) \gets (m_{i-1}^{-1} \bmod m_i,\ldots, m_{i-1}^{-1} \bmod m_k)$
        \State $(y_i,\ldots,y_k) \gets (y_i,\ldots,y_k)\cdot(t_i,\ldots,t_k)$
    \EndFor
    \State $(r_1,\ldots,r_k) \gets (m_1,\ldots,m_k)$ \MyComment{Associated MRNS base}
\end{algorithmic}
\end{algorithm}

\section{Silenzio}
We introduce our approach for non-interactive outsourcing of MLP training without any non-collusion assumptions.

\begin{table}
    \centering
    \caption{Notations}
    \label{tab:notation}
    \begin{tabular}{c l}
        \toprule
        \textbf{Notation} & \textbf{Description} \\
        \midrule
        $\mathcal{U}_b$ & $\mathbb{Z} \cap [0,2^b]$\\
        $\mathcal{I}_b$ & $\mathbb{Z} \cap [-2^{b-1},2^{b-1}-1]$\\
        $\mathcal{J}_b$ & $\mathcal{U}_b$ values representing a number in RNS\\
        $\mathcal{K}_b$ & $\mathcal{U}_b$ values representing a number in MRNS\\
        $\mathbf{m}$    & RNS base: vector of modules\\
        $w$             & Bitwidth of the RNS modules\\
        $\alpha$        & Signed bitwidth of the model parameters\\
        $\beta$         & Signed bitwidth of the inferece/training data\\
        $\Gamma$, $\gamma$ & Signed / unsigned output bitwidth\\
        $x$             & Cap value of $\operatorname{ReLU}_x$\\
        
        \bottomrule
    \end{tabular}
\end{table}

\subsection{Threat Model}
\label{sec:threat_model}
We assume a scenario where the client outsources the training of an MLP to an untrusted cloud-service-provider under a semi-honest, or honest-but-curious, adversary model. The provider strictly follows the protocol but may attempt to learn information from the data it processes. All training data and model parameters are encrypted using TFHE configured to achieve 128\,bit security. For reusing encrypted training data and the successive updating of the weights during training, we rely on TFHE's security guarantees enabling the secure reuse of encrypted inputs. For our implementation of \schemename, we leveraged Zama's Concrete Library with the default parameters, which means the implementation is secure in the IND-CPA security model. Concretely, the cloud customer, aka model owner, should not share the results of the outsourced computation (e.g., the trained model or inference results) in cleartext with third parties, as this could leak the secret key. The confidentiality of the system relies on the assumption that the client securely stores the secret key and never discloses it to any third party, thereby preventing the provider from decrypting any sensitive information. This model is designed solely to preserve confidentiality and does not extend to protect against active or integrity-based attacks, as it does not include mechanisms for detecting protocol deviations. Although our threat model is similar to that of FHESGD \cite{DBLP:conf/cvpr/NandakumarRPH19} and Glyph \cite{DBLP:conf/nips/LouFF020}, which also operates under a semi-honest assumption, our approach significantly strengthens the security guarantee by employing a realistic 128\,bit security level in contrast to FHESGD's and Glyph’s 80\,bit setting.

\subsection{Setup}
\begin{figure}
	\centering
	\includegraphics[width=\linewidth]{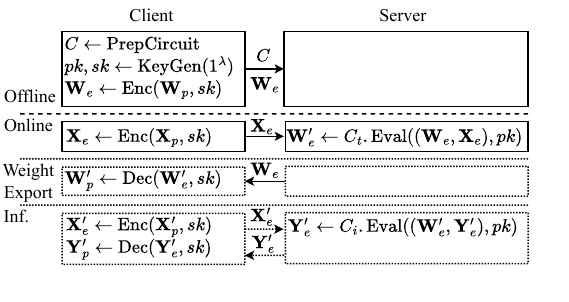}
	\caption{\schemename's set up divided into an input-independent offline and an online phase. Here, we denote plaintexts and ciphertexts with a subscript $p$ respective $e$ and indicate the evaluation of a circuit through $C.\operatorname{Eval}()$.}
	\label{fig:set-up}
\end{figure}

 As visualized in \cref{fig:set-up}, \schemename{} is set up in two stages, we call offline and online phase. In the input-independent \emph{offline phase}, the client prepares the public evaluation and secret en-/decryption keys. Further, they prepare the TFHE-protected circuit, including the randomly initialized and encrypted weights, and transmit them to the remote cloud-service-provider. In the \emph{online phase}, the client sends their encrypted dataset to the server, and the server performs the protected training. After the training is finished, the client can either use the encrypted weights on the server or request the server to send the trained weights back for decryption and local inference usage. For low-power devices like smart sensors in the IoT context, it's also thinkable that the client sends new training data points continuously to unburden local storage resources and keep the outsourced MLP model through continuous training up-to-date. Note that in \schemename{} and its evaluation, we concentrate on the training and do not consider the inference phase. To accelerate the inference after the training phase, it might be interesting to explore ways of combining the trained weights of \schemename{} with a more inference-focused approach like REDSec~\cite{DBLP:conf/ndss/FolkertsGT23}.

\newcommand*\rot[1]{\hbox to1em{\hss\rotatebox[origin=br]{-50}{#1}}}

\begin{table}[t!]
  \centering
  \caption{Recent neural network frameworks in the non-interactive outsourcing setting sorted by target workload. Some framework names are made up for easier referencing. LFE in fine-tuning means the client has to perform feature extraction with a public pre-trained model locally and then sends these features to the server to train a new model. $^*$According to the authors, BlindTuner will be open sourced upon acceptance. HETAL communicates the model output of a validation set after each epoch to provide a client-driven early-stopping feature, but apart from that feature, HETAL is also non-interactive.}
  \label{tab:featuresetcomparison}
  \setlength{\tabcolsep}{2pt}
  \scalebox{1.0}{
  \begin{threeparttable}
  \begin{tabular}{llcccc}
    \toprule
    & Name & Scheme & \rot{Enc. Weights} & \rot{Sec. Level} & \rot{Open Source}\\
    \midrule

    \multirow{12}{*}{\rotatebox[origin=c]{90}{\textbf{Inference}}}
    & CryptoNets \cite{DBLP:conf/icml/Gilad-BachrachD16} & YASHE & $\Circle$ & 80 & $\CIRCLE$ \\ %
    & Faster CryptoNets\cite{DBLP:journals/corr/abs-1811-09953} & FV-RNS & $\Circle$ & 128 & $\Circle$\\ %
    & FHE-DiNN\cite{DBLP:conf/crypto/BourseMMP18} & TFHE & $\Circle$ & 80 & $\CIRCLE$\\ %
    & LoLa\cite{DBLP:conf/icml/BrutzkusGE19} & BFV & $\Circle$ & 128 & $\CIRCLE$\\ %
    & MPCNN\cite{DBLP:conf/icml/LeeLLK0NC22} & CKKS & $\Circle$ & 128 & $\Circle$ \\ %
    & ResFHE\cite{DBLP:journals/access/LeeKLCEDLLYKN22} & CKKS & $\Circle$ & 111.6 & $\Circle$ \\ %
    & REDsec\cite{DBLP:conf/ndss/FolkertsGT23} & TFHE & $\CIRCLE$ & 128 & $\CIRCLE$ \\ %
    & AutoFHE\cite{DBLP:conf/uss/AoB24} & CKKS & $\Circle$ & 128 & $\CIRCLE$\\ %
    & DaCapo\cite{DBLP:conf/uss/Cheon00LJKL024} & CKKS & $\Circle$ & 128 & $\CIRCLE$ \\ %
    & NeuJeans\cite{DBLP:conf/ccs/JuP0KKCA24} & CKKS & $\Circle$ & 128 & $\Circle$ \\ %
    & Nexus\cite{DBLP:journals/iacr/ZhangLYWCHRY24} & CKKS & $\CIRCLE$ & 128 & $\CIRCLE$ \\ %
    & LowMemInf\cite{DBLP:journals/ijns/RovidaL24} & CKKS & $\Circle$ & 128 & $\CIRCLE$ \\ %
    & LOHEN\cite{nam2025lohen} & CKKS, TFHE & $\Circle$ & 128 & $\Circle$ \\ %
    & FHE-Neuron\cite{DBLP:journals/popets/KuLHCHTC25} & TFHE & $\Circle$ & 128 & $\Circle$ \\ %

    \midrule
    \multirow{3}{*}{\rotatebox[origin=c]{90}{\parbox{1cm}{\centering\textbf{Fine-t.}}}}
    & PrivGD\cite{DBLP:conf/provsec/JinRA20} (LFE) & CKKS & $\CIRCLE$ & 80 & $\Circle$\\ %
    & HETAL\cite{DBLP:conf/icml/LeeLKSL23} (LFE) & CKKS & $\CIRCLE$ & 128 & $\CIRCLE$\\ %
    & BlindTuner\cite{DBLP:journals/corr/abs-2402-09059} (LFE) & CKKS & $\CIRCLE$ & 128 & $*$ \\ %
    & Glyph\cite{DBLP:conf/nips/LouFF020} & TFHE, BGV & $\CIRCLE$ & 80 & $\Circle$\\ %

    \midrule
    \multirow{3}{*}{\rotatebox[origin=c]{90}{\textbf{Train.}}}
    & FHESGD \cite{DBLP:conf/cvpr/NandakumarRPH19} & BGV & $\CIRCLE$ & $\sim80$ & $\Circle$\\ %
    & Glyph\cite{DBLP:conf/nips/LouFF020} & TFHE, BGV & $\CIRCLE$ & 80 & $\Circle$\\ %
    & \schemename & TFHE & $\CIRCLE$ & 128 & \CIRCLE\\

    \bottomrule
  \end{tabular}
  \begin{tablenotes}
    \centering
    \item $\CIRCLE \text{ Support}$ $\Circle \text{ No Support}$
  \end{tablenotes}
  \end{threeparttable}}
\end{table}

\subsection{Encoding \& Bitwidths}
Through NITI, \etal{Wang} \cite{DBLP:journals/tpds/WangRLS22} demonstrated the feasibility of training NNs using 8\,bit weights while down-scaling activations, gradients, and the errors also equal to or below 8\,bits. Previous works like \cite{DBLP:conf/iclr/ZhuHMD17, DBLP:conf/eccv/RastegariORF16, DBLP:journals/corr/CourbariauxB16, DBLP:conf/cvpr/JacobKCZTHAK18, DBLP:journals/corr/ZhouNZWWZ16, DBLP:conf/nips/BannerHHS18, DBLP:conf/iclr/WuLCS18, DBLP:conf/ijcnn/ChenHZX17} showed the effectiveness of using integer weights below 8\,bits for training, but their activations, errors, or gradients are much larger or quantized from 32\,bit floating-point values, making them unsuitable for our integer-only-based training approach. Following NITI, \schemename{} leverages $\alpha\leq8$\,bit signed integer weights for effective training. To make \schemename's whole end-to-end training pass low-bitwidth compatible, we additionally propose a new low-bitwidth gradient computation for the cross-entropy loss, further reducing the bitwidth requirements compared to NITI.

State-of-the-art FHE implementations, like the Concrete library, are most effective for bitwidths smaller or equal to 8\,bits.
To allow effective training while maintaining good runtime performance, \schemename{} only computes on tiny integer values for the entire computation, never exceeding 8\,bits, and only using RNS representations where needed. To constrain the bitwidth of the used values below or equal to 8\,bits and enable \schemename's optimized operations, we leverage the RNS bases shown in \cref{tab:rns-bases} for the evaluation. Note that \schemename{} also supports larger bitwidths as shown in \cref{tab:rns_bases_5bit} and explained in \cref{sec:low_bitwidth_matmul}.

To summarize, we follow previous work from the hardware community \cite{DBLP:journals/tpds/WangRLS22} showing the effectiveness of training using low-bitwidth integer weights and errors and exploit it to boost the runtime performance of FHE-protected training. \schemename{} works with integer in the finite ring \(\ZZ_{M_k}\). To represent signed integer, positive values are assigned to the lower half of \(\ZZ_{M_k}\), and negative values to the upper half. The mapping \((0, 1, 2, -2, -1) \mapsto (0, 1, 2, 3, 4)\) illustrates the encoding.

\begin{table}
\centering
\caption{RNS bases used in \schemename.}
\label{tab:rns-bases}
\begin{tabular}{llc}
\toprule
$k$ & RNS Base & Max. Bitwidth\\\midrule
2 & 15, 14 & 7.72\\
3 & 13, 15, 14 & 11.41\\
4 & 11, 13, 15, 14 & 14.87\\
5 & 7, 11, 13, 15, 8 & 16.87\\
6 & 5, 7, 9, 11, 13, 8 & 18.45\\
\bottomrule
\end{tabular}
\end{table}

\subsection{Training Scheme}
\label{sec:training_scheme}
\begin{figure*}
	\centering
	\includegraphics[width=0.7\linewidth]{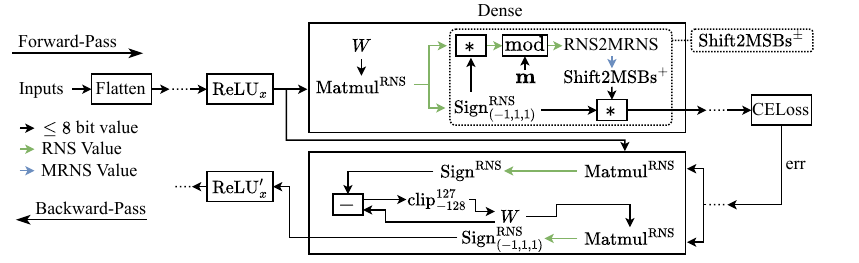}
	\caption{High-level overview of the back- and forward-pass in \schemename{}.}
	\label{fig:overview}
\end{figure*}

In the following, we introduce each of \schemename's components in detail to finally outline the whole end-to-end training approach. For simplicity, we leverage NumPy's broadcasting and indexing rules \cite{harris2020array} in the descriptions of the algorithms. Particularly, the $\operatorname{expandDims}(tensor, axis)$ function has the same syntax and semantics as in NumPy, changing the dimensions of a tensor without changing the data. Additionally, we leverage the FHE components described in \cref{sec:fhe} and Concrete's efficient implementations for modular reduction $\bmod$, $\max$, $\operatorname{ReLU}$ and bit-extraction $\operatorname{extractBits}(input, indices)$ as black boxes. We denote our custom lookup tables for use with Concrete's PBS as $\operatorname{lookupComp}()$ and multivariate functions with two encrypted inputs as $\operatorname{multivariateLookupComp}()$. \Cref{tab:notation} introduces the used notations. \cref{fig:overview} shows the interplay of \schemename's components and the used number systems from a high-level perspective. 

\subsubsection{$\text{ReLU}_x$}
To keep the non-linear activations of our models in manageable value ranges, we follow the approach of \etal{Krizhevsky}~\cite{krizhevsky2010convolutional}, cap the ReLU function and call it now 
\begin{equation*}
\operatorname{ReLU}_x(a)=\min(\max(a, 0), x) \text{ with } x \geq 0.
\end{equation*}
While \etal{Krizhevsky} fixed $x=6$, we evaluated \schemename{} for a wide range of cap values $x\in [0,127]$ and found $x=14$ to generally be a suitable threshold for \schemename's low-bitwidth integer training approach.

\subsubsection{Low-bitwidth Matrix Multiplication}
\label{sec:low_bitwidth_matmul}
As shown in \cref{fig:overview}, the matrix multiplication $\operatorname{Matmul}^{\text{RNS}}$ in \schemename{} gets up to 8\,bit matrices on the forward- as well as the backward-pass as inputs. To keep all FHE-processed values $\leq$ 8\,bit, we perform a conversion of both inputs into an RNS system (see also \cref{lst:matmul}).
Since the used RNS bases consist of 4\,bit moduli, all residues are $\leq$ 4\,bit and we can compute the partial products in step 2 without the intermediate results overflowing 8\,bit.
Finally, the summation of the matrix multiplication is performed in the last step. We split the summation into modularly reduced sub-sums of maximal 15 summands to not overflow our limit of 8\,bit intermediate results. Finally, all these sub-sums are summed up (modulo RNS base) to form the final result. Note that we can select the used RNS base individually per $\operatorname{Matmul}$ operation based on the required bitwidth. A computation including multiple $\operatorname{Matmul}$ operations of different input bitwidths and dimensions, like MLP training, greatly benefits in terms of computational efficiency from a heterogeneous set of RNS bases.

As part of \schemename{}, we additionally developed a high-resolution variant, $\operatorname{MatmulHighRes}^{\text{RNS}}$, which supports much larger bitwidths for the output of the matrix multiplication and may enable future work to train larger neural networks like, e.g., CNNs. $\operatorname{MatmulHighRes}^{\text{RNS}}$ leverages a new Karatsuba-inspired multiplication routine enabling the usage of 5\,bit moduli in the RNS base and is fully compatible with all other components of \schemename. See \cref{sec:low_bw_high_res_matmul} for a detailed description, including a numerical step-by-step example in \cref{fig:matmul_example}.
\begin{algorithm}%
\caption{$\operatorname{Matmul}^{\text{RNS}}$}
\label{lst:matmul}
\begin{algorithmic}
    \Require $\mathbf{X} \in \mathcal{I}_8^{a \times b}$, $\mathbf{W} \in \mathcal{I}_8^{b \times c}$, $\mathbf{m} \in \mathcal{U}_4^k$ \MyComment{Inputs, RNS base}
    \Ensure $\mathbf{Y} \in \mathcal{J}_4^{k \times a \times b}$ \MyComment{Output in RNS representation}

    \LineComment{1. Step: Convert inputs to RNS representation}
    \State $\mathbf{m}' \gets \operatorname{expandDims}(m, [1, 2])$
    \State $\mathbf{X}' \gets \mathbf{X} \bmod \mathbf{m}'$ \MyComment{Shape: $k\times a \times b$}
    \State $\mathbf{W}' \gets \mathbf{W} \bmod \mathbf{m}'$ \MyComment{$k \times b \times c$}

    \LineComment{2. Step: Compute partial products}
    \State $\mathbf{X}' \gets \operatorname{expandDims}(\mathbf{X}', 3)$ \MyComment{$k \times a \times b \times 1$}
    \State $\mathbf{W}' \gets \operatorname{expandDims}(\mathbf{W}', 1)$ \MyComment{$k \times 1 \times b \times c$}
    \State $\mathbf{Y}^* \gets \mathbf{X}' \mathbf{W}' \bmod \operatorname{expandDims}(m, [1, 2, 3])$\MyComment{$k \times a \times b \times c$}

    \LineComment{3. Step: Block-wise summation}
    \State $n \gets \min(15,b)$
    \State $\mathbf{Y} \gets \operatorname{sum}(\mathbf{Y}^*[:, :, :n, :], axis=2) \bmod \mathbf{m}'$
    \For{$i \gets n,\; i \le b,\; i \gets i+n$}
        \State $j \gets \min(i+n, b)$
        \State $\mathbf{Y}' \gets \operatorname{sum}(\mathbf{Y}[:,:,i:j,:], axis=2)$
        \State $\mathbf{Y} \gets \mathbf{Y} + \mathbf{Y}' \bmod \mathbf{m}'$\MyComment{$k \times a \times c$}
    \EndFor
    
\end{algorithmic}
\end{algorithm}

\subsubsection{Sign Determination in RNS}
\label{sec:sign_determination}
To extract the sign information of a value represented in an RNS, we start with a simple and widely used approach. We first convert the number to the associated MRNS using our vectorized implementation of the RNS2MRNS conversion shown in \cref{lst:rns_to_mrs_conversions}. Subsequently, we exploit that our encoding divides the finite ring $\ZZ_{M_k}$ of all possible values into the lower half for positive and the upper half for negative values. Now, we can extract the sign information using a single table lookup on the most significant radix position: $x_k\geq r_k/2$. To achieve full correctness,  the sign extraction requires the most significant radix to be even. The most significant modulus of our RNS bases is always chosen to be even (see \cref{tab:rns-bases}) and as the RNS2MRNS conversion results in an MRNS representation of an associated MRNS base, we always meet this requirement and achieve correct sign extraction.

Additionally, \schemename{} can check for zero-equality. Therefore, we add the values of the MRNS representation along all radices and compare to zero. As our RNS bases, and so our MRNS bases never contain more than six 4 (or 5)\,bit moduli, their sum never exceeds 8\,bit.
Note that we could similarly construct the zero-equality check based on the RNS representation, but as we cannot jump (in the algorithm) based on encrypted values, there is no benefit to doing so.
Leveraging the correct sign extraction and the zero-equality check, \schemename{} supports sign gadgets with the following semantics:
\begin{equation*}
    \operatorname{Sign}_{(n,z,p)}^{\text{RNS}}(x):=
    \begin{cases}
    n,\text{ for } x < 0\\
    z,\text{ for } x= 0\\
    p,\text{ for } x>0.\\
    \end{cases}
\end{equation*}

\subsubsection{$\text{SHIFT2MSBs}^+$}
One of \schemename's key features is a new block-scaling gadget that allows us to approximately shift an input matrix given in RNS representation to its user-specified $\gamma$ most significant bits. Therefore, the gadget 
approximates the maximum bitwidth present in the input and performs a shift operation based on this maximum.
The gadget is central to keeping all FHE-processed values $\le$ 8\,bit while performing a high-resolution forward-pass and enabling a fruitful backward-pass that allows not only protected inference, but training resulting in accurate MLP models.

We start by describing $\text{SHIFT2MSBs}^+$ which allows shifting positive RNS represented matrices, and subsequently introduce $\text{SHIFT2MSBs}^\pm$ which also considers negative values. \Cref{lst:Shift2MSBs_p} shows $\text{SHIFT2MSBs}^+$ and \cref{tab:shift2msb} illustrates the algorithm exemplary step-by-step (without the RNS-to-MRNS conversion step). The algorithm gets a positive valued matrix in RNS representation $\mathbf{X}$ and the user-specified unsigned output bitwidth $\gamma$ as inputs. Note the unsigned output bitwidth $\gamma$ is just one bit less than the signed output bitwidth $\Gamma$. In the first step, the input matrix is converted from an RNS to the associated MRNS representation. From now on, the main idea of the gadget is to view the MRNS values as if they were chunked binary numbers, where each chunk is given by a radix-digit (see the second column in the example \cref{tab:shift2msb}).

Based on the MRNS representation, we approximately extract the position of the highest set bit, $maxBit$, in step 2. Step 3 computes the amount of shift needed per radix-digit $digitShift$ to achieve an output bitwidth $\gamma$. First we check, whether any shift is required by comparing $maxBit$ to the radix bitwidth $w$, as we only want to shift down but never up. Shifting up would not provide any additional information to the training process. Then we compute the shift required per radix digit $digitShift$ based on $maxBit$, the required output bitwidth $\gamma$ and the maximum bitwidth per radix-digit $maxBW$. Finally, we compute the amount of required left and right shift based on $digitShift$.

In the fourth step, we just apply the left and right shifts per radix-digit on the MRNS representation. To get the final result, we sum over all radix-digits in step 5. Notice that step 4 and step 5 inherently convert the matrix from an MRNS to the classical decimal representation. We approximately compute the effectively performed shift amount $shift$ in the last step of the algorithm.

As the radices in our MRNS representations are not all powers of two, the approximation error of the gadget stems from the underlying idea of viewing the radix-digits as if they would make up chunks of a binary number. The error has two sources: first, the approximative determination of the highest set bit, and second, the per radix-digit bit-shifts together with the final digit summation, which do not consider the semantics of the MRNS representation. We analyze the approximation error as part of our evaluation in the practical application of MLP training.
\begin{algorithm}[t!]
\caption{$\text{Shift2MSBs}^+$}
\label{lst:Shift2MSBs_p}
\begin{algorithmic}
    \Require $\mathbf{X} \in \mathcal{J}_4^{k \times a \times b}$, $\mathbf{m} \in \mathcal{U}_4^k$, $w$, $\gamma$ \MyComment{Inputs, RNS base, RNS modul bitwidth, Unsigned output bitwidth}
    \Ensure $\mathbf{Y} \in \mathcal{U}_\gamma^{a \times b}$, $shift$ \MyComment{Output, Shiftamount}
    
    \LineComment{1. Step: RNS-to-MRNS conversion}
    \State $\mathbf{X}', \_ \gets \text{RNS2MRNS}(\mathbf{X}, \mathbf{m})$

    \LineComment{2. Step: Extract the position of the highest set bit (1-indexed)}
    \State $maxBit \gets 0$
    \For{$i \gets 0, i<k,i\gets i+1$}
        \State $\mathbf{u} \gets \operatorname{lookupComp}(\lceil\log_2(\mathbf{X}'[i]+1)\rceil+i\cdot w)$
        \State $maxBit \gets \max(\max(\mathbf{u}), maxBit)$
    \EndFor

    \LineComment{3. Step: Compute amount of digitShift}
    \State $anyShift \gets maxBit > w$ \MyComment{Only shift down, never up}

    \For{$i \gets 0, i<k, i\gets i+1$} \MyComment{Can be pre-computed}
        \State $maxBW[i] \gets w \cdot i$
    \EndFor
    \State $digitShift \gets (\gamma - (maxBit - maxBW))\cdot anyShift$
    \State $lshift \gets \min(\max(digitShift, 0), \gamma - 1)$
    \State $rshift \gets \operatorname{ReLU}(-digitShift)$

    \LineComment{4. Step: Perform Shift}
    \State $\mathbf{Y}^* \gets \mathbf{X}' \ll \operatorname{reshape}(lshift, [k,1,1])$
    \State $\mathbf{Y}^* \gets \mathbf{Y}^* \gg \operatorname{reshape}(rshift, [k,1,1])$

    \LineComment{5. Step: Reduce with final summation}
    \State $\mathbf{Y} \gets \operatorname{sum}(\mathbf{Y}^*, axis=0)$

    \LineComment{6. Step: Compute amount of shift}
    \State $shift \gets maxBit-\gamma$
\end{algorithmic}
\end{algorithm}

\begin{table}
\centering
\caption{Example computation of $\text{Shift2MSBs}^+$ with an input of size 3 and a batch of a single sample. Unsigned output bitwidth $\gamma=5$ and RNS modul bitwidth $w=4$. Shift results are underlined and for brevity, the RNS-to-MRNS conversion is omitted. }
\label{tab:shift2msb}
\setlength{\tabcolsep}{4pt}
\begin{tabular}{l|c|c|c}
\toprule
MRNS Base & $r_2=14$ & $r_1=15$ & $r_0=13$ \\
& 1110 & 1111 & 1101\\ \midrule
$\mathbf{X}'[:,0,0]$ & $00\underline{11}$ & $\underline{001}0$ & $0000$\\
$\mathbf{X}'[:,0,1]$ & $00\underline{01}$ & $\underline{110}0$ & $0010$\\
$\mathbf{X}'[:,0,2]$ & $00\underline{00}$ & $\underline{000}1$ & $0110$\\ \midrule
$anyShift$ & \multicolumn{3}{c}{$(maxBit > w) = (10 > 4) \implies \text{True}$} \\ \midrule
$digitShift$    & $5 - (10 - 8)$ & $5 - (10 - 4)$ & $5 - (10 - 0)$ \\
$lShift$   & 3          & 0          & 0          \\
$rshift$  & 0      & 1      & 5      \\ \midrule\midrule
$shifted$ & $\mathbf{X}'[:,0,0]$ & $\mathbf{X}'[:,0,1]$ & $\mathbf{X}'[:,0,1]$ \\
& $\underline{11}000$ & $\underline{01}000$ & $\underline{00}000$\\
& $00\underline{001}$ & $00\underline{110}$ & $00\underline{000}$\\
& $00000$ & $00000$ & $00000$\\\midrule
$\mathbf{Y}$          & $\underline{11001}$      & $\underline{01110}$      & $\underline{00000}$      \\\midrule
$shift$ & \multicolumn{3}{c}{$maxBit-\gamma=10-5=5$}\\
\bottomrule
\end{tabular}
\end{table}

\subsubsection{$\text{SHIFT2MSBs}^\pm$}
To allow the shifting of negative numbers in RNS representation, we introduce $\text{SHIFT2MSBs}^\pm$ shown in \cref{lst:Shift2MSBs_pm}. The main idea is that shifting in the negative domain is congruent to shifting in the positive domain and that our encoding allows for efficient computation of absolute values given a number in RNS representation.

The algorithm gets an input matrix $\mathbf{X}$ in RNS representation and the signed output bitwidth $\Gamma$ as inputs. First, we extract the sign information $s$ of the input using the approach introduced in \cref{sec:sign_determination}. Using the sign information, we compute the absolute values $\mathbf{X}^+$ of the input in RNS representation. Afterward, we use our $\text{SHIFT2MSBs}^+$ algorithm to shift the absolute values to the $\Gamma-1$ most significant bits. In the last step, we reapply the sign information, extracted in the beginning, to the shifted absolute values $\mathbf{Y}^+$ resulting in the final output $\mathbf{Y}$.

\begin{algorithm}
\caption{$\text{Shift2MSBs}^{\pm}$}
\label{lst:Shift2MSBs_pm}
\begin{algorithmic}
    \Require $\mathbf{X} \in \mathcal{J}_4^{k \times a \times b}$, $\mathbf{m} \in \mathcal{U}_4^k$, $w$, $\Gamma$ \MyComment{Inputs, RNS base, RNS modul bitwidth, Signed output bitwidth}
    \Ensure $\mathbf{Y} \in \mathcal{I}_\Gamma^{a \times b}$, $shift$ \MyComment{Output, Shiftamount}

    \State $s \gets \operatorname{Sign}_{(-1,1,1)}^{\text{RNS}}(\mathbf{X})$ \MyComment{Extract sign}
    \State $\mathbf{X}^+ \gets s \cdot \mathbf{X} \bmod \operatorname{reshape}(\mathbf{m}, [k,n,n])$ \MyComment{Extract abs. values}

    \State $\mathbf{Y}^+, shift \gets \text{Shift2MSBs}^+(\mathbf{X}^+, \mathbf{m}, w, \Gamma-1)$

    \State $\mathbf{Y} \gets s \cdot \mathbf{Y}^+$ \MyComment{Add sign back to result}
    
\end{algorithmic}
\end{algorithm}

\subsubsection{Integer Cross-Entropy Loss Derivative}
\label{sec:integer_cross_entropy_loss_derivative}
Given the logits $a_i$ at the output of the final layer of an MLP, the corresponding softmax distribution is defined as $\hat{y}_i = e^{a_i}/\sum_i e^{a_i}$. We can now compute the error based on the cross-entropy loss, which is defined as $L = -\sum_i y_i \ln(\hat{y}_i)$ with $\mathbf{y}$ being the target probabilities given through the labels in the training data. The error is computed as the partial derivatives of the loss with respect to the MLP outputs 
\begin{equation*}
    \partial L / \partial a_i = \hat{y}_i - y_i.    
\end{equation*}

\begin{figure}
	\centering
	\includegraphics[width=\linewidth]{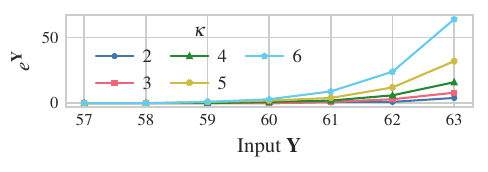}
	\caption{Input-output behavior of our approximated exponential function $\operatorname{exp}(\mathbf{Y}) =\operatorname{round}((e^{\mathbf{Y}})/e^{2^6-1}\cdot2^\kappa)$ on signed 7\,bit inputs for various approximation level $\kappa$.}
	\label{fig:exp_approx}
\end{figure}

As shown in \cref{lst:IntCELossDeriv} \schemename{} approximates the error computation, depending solely on integer computations. The algorithm expects the logits on the output of the last layer in the MLP $\mathbf{\hat{Y}}$, the corresponding one-hot encoded labels of the training dataset $\mathbf{Y}$ and an approximation level $\kappa$ for the exponential function. First, we approximate the exponential function using $\lceil\log_2(2^\kappa+1)\rceil$ bit integer through a table lookup. \Cref{fig:exp_approx} shows a plot of the approximation for different approximation levels $\kappa$. As we sum these approximations in the next step and still hold the requirement of not exceeding 8\,bit intermediate values, the algorithm is limited to MLPs with up to $\log_2(o\cdot 2^\kappa + 1) \leq 8 \implies o \leq 255/2^\kappa$ output neurons. To compute the error, we leverage a multivariate table lookup with two encrypted inputs and subtract the target ``probabilities'' given through the training labels. Finally, for training MLPs, \schemename{} just extracts the sign of the error and leverages it for the backpropagation algorithm. Compared to the derivative computation in  NITI, \schemename{} depends on much smaller integer, never exceeding 8\,bit. %
\begin{algorithm}[t!]
\caption{IntCELossDeriv}
\label{lst:IntCELossDeriv}
\begin{algorithmic}
    \Require $\mathbf{\hat{Y}} \in \mathcal{I}_\Gamma^{a \times b}$, $\mathbf{Y} \in \mathcal{U}_1^{a \times b}$, $\text{with } b\leq 16$, $\kappa$ \MyComment{Logits, One-hot encoded label, exp. approx. level}
    \Ensure $\mathbf{E} \in \mathcal{I}_{\lceil\log_2(b\cdot2^\kappa+1)\rceil}^{a \times b}$ \MyComment{Error}

    \LineComment{$round(x)$ means rounding $x$ to the next integer. If $x$ is exactly between two integer we round to the nearest even integer.}
    \State $\mathbf{E} \gets \operatorname{ReLU}(\mathbf{\hat{Y}})$
    \State $\mathbf{E} \gets \operatorname{lookupComp}(\operatorname{round}((e^{\mathbf{\hat{Y}}})/e^{2^\gamma-1}\cdot2^\kappa))$
    \State $\mathbf{S} \gets \operatorname{reshape}(\operatorname{sum}(\mathbf{E}, axis=1), [a, 1])$

    \State $\mathbf{E} \gets \operatorname{multivariateLookupComp}(\operatorname{round(\mathbf{E}\cdot2^\kappa+1/\mathbf{S}+1)})$

    \State $\mathbf{E} \gets \mathbf{E} - \mathbf{Y}\cdot \operatorname{reshape}(\operatorname{sum}(\mathbf{E}, axis=1),[a,1])$
    \State $\mathbf{E} \gets \operatorname{lookupComp((\mathbf{E}>0)-(\mathbf{E}<0))}$ \MyComment{Extract sign}
\end{algorithmic}
\end{algorithm}

\subsubsection{End-to-End Training}

\Cref{lst:EndToEndTraining} shows \schemename's end-to-end training approach for a single data batch and for brevity without activation functions. \schemename{} leverages stochastic gradient descent based optimization for effective training. Training over multiple data batches and epochs can be performed through repeated execution of the training pass while continuously updating the same encrypted weights $\mathbf{W}$. Before every matrix multiplication, we perform an unencrypted table lookup using $\operatorname{getRNSBase}()$ on \cref{tab:rns-bases} to set the used RNS base for the required operand resolution, determined by the input dimensions and bitwidths.

On the forward-pass, we leverage $\text{Shift2MSBs}^\pm$ to reduce the bitwidth of the matrix multiplication results back to $\Gamma$\,bits while simultaneously converting it from the RNS to the decimal number system. While $\text{Shift2MSBs}^\pm$ also provides the amount of shift performed, \schemename{} does not use it for the training. As demonstrated by \etal{Wang} \cite{DBLP:journals/tpds/WangRLS22} in an unprotected setting, the amount of shift could be used in future work to construct schemes to train ML models which require the amount of shift, also known as scaling exponent in fixed-point arithmetic, for effective training.

We instantiate the back-propagation of the error using our approximated cross-entropy loss derivation. During MLP-Training \schemename{} just leverages the sign of the error for backpropagation. Similar to previous work \cite{DBLP:conf/icnn/RiedmillerB93, DBLP:journals/tpds/WangRLS22} we perform the weight update based on the sign of the weight gradients. We leverage $\operatorname{Sign}^{\text{RNS}}$ and extract the sign information directly from the outputs of the $\operatorname{Matmul}^{\text{RNS}}$ computations, omitting additional $\text{Shift2MSBs}^\pm$ operations for downshifting and number-system conversion. Finally, to limit the weights to 8\,bit during the weight update and prevent intermediate results greater than 8\,bit, we leverage a multivariate table lookup. If required, a bias value could be introduced through the bias-trick by appending a constant 1-entry to the input and adding a bias column to the weight matrix.
\begin{algorithm}[t!]
\caption{End-to-End Training (Single Batch)}
\label{lst:EndToEndTraining}
\begin{algorithmic}
    \Require $\mathbf{A}_0 \in \mathcal{I}_\beta^{a_0 \times b_0}$, $\mathbf{Y} \in \mathcal{U}_1^{a \times b}$, $\alpha$, $\beta$, $\Gamma$, $x$ \MyComment{Data batch, One-hot enc. labels, model bitwidth, input bitwidth, signed output bitwidth, $ReLU_x$ cap value}
    \Ensure $\mathbf{W}_l \in \mathcal{I}_\alpha^{c_l \times b_l} \text{ for all } l \in [1,L]$ \MyComment{MLP weights}
    \LineComment{We train an MLP of $L$ weight layers. For brevity, we omit the activation functions.}
    
    \LineComment{Forward-Pass}
    \For{$l \gets 1, l\leq L, l\gets l+1$}
        \If{$l=1$}
            \State $\mathbf{m}\gets \operatorname{getRNSBase}(\log_2(b_{l-1}\cdot 2^{\beta}\cdot2^{\alpha}+1))$
        \Else
            \State $\mathbf{m}\gets \operatorname{getRNSBase}(\log_2(b_{l-1}\cdot x\cdot2^{\alpha}+1))$
        \EndIf
        \State $\mathbf{A}_l\gets \operatorname{Matmul}^{\text{RNS}}(\mathbf{A}_{l-1}, \mathbf{W}_l^\top, \mathbf{m})$ \MyComment{Layer l activation}
        \State $\mathbf{A}_l\gets \text{Shift2MSBs}^\pm(\mathbf{A}_l, \mathbf{m}, 4, \Gamma)$ \MyComment{4\,bit RNS moduls}
    \EndFor

    \LineComment{Backward-Pass}
    \State $\mathbf{E}\gets \operatorname{IntCELossDeriv}(\mathbf{A}_l, \mathbf{Y})$
    \For{$l \gets L, l\geq 1,l\gets l-1$}
        \If{$l>1$}
            \State $\mathbf{m}\gets \operatorname{getRNSBase}(\log_2(a_{l-1}\cdot2\cdot x+1))$
            \State $\mathbf{G}\gets \operatorname{Matmul}^{\text{RNS}}(\mathbf{E}^\top, \mathbf{A}_{l-1}, \mathbf{m})$
            \State $\mathbf{m}\gets \operatorname{getRNSBase}(\log_2(c_l\cdot2^\alpha+1))$
            \State $\mathbf{E}\gets \operatorname{Matmul}^{\text{RNS}}(\mathbf{E}, \mathbf{W}_l, \mathbf{m})$
            \State $\mathbf{E}\gets \operatorname{Sign}_{(-1,0,1)}^{\text{RNS}}(\mathbf{E})$ \MyComment{Sign for error propagation}
        \Else
            \State $\mathbf{m}\gets \operatorname{getRNSBase}(\log_2(a_{l-1}\cdot 2^\beta+1))$
            \State $\mathbf{G}\gets \operatorname{Matmul}^{\text{RNS}}(\mathbf{E}^\top, \mathbf{A}_{l-1}, \mathbf{m})$
        \EndIf
        \State $\mathbf{G}\gets \operatorname{Sign}_{(-1,0,1)}^{\text{RNS}}(\mathbf{G})$ \MyComment{Sign for weight update}
        \State $\mathbf{W}\gets \operatorname{multivariateLookupComp}(\operatorname{clip}_{-128}^{127}(\mathbf{W}-\mathbf{G}))$
    \EndFor
\end{algorithmic}
\end{algorithm}

\section{Implementation}
We describe our implementation, used to evaluate the individual components and \schemename's end-to-end performance.

\subsection{Setup}
We used a server running Ubuntu 24.04 equipped with 2x AMD EPYC 9534 and the following software: Python 3.10.13, Concrete-Python 2.11.0, Jax 0.6.2, PyTorch 2.8.0 and scikit-learn 1.7.1. We leverage Concrete-Python to compile \schemename's components into an efficient TFHE program. To speed up the development and evaluation, we also implemented a fast clear-text clone based on Jax-accelerated NumPy that performs the same approximations as the FHE-based components. To compare the prediction-performance of the MLPs trained with \schemename{}, we used standard PyTorch to construct models of the same architecture as a baseline.

\subsection{Circuit Construction}
Congruent to classical NN frameworks like TensorFlow and PyTorch and, as depicted in \cref{fig:overview}, \schemename{} models MLPs in components of fully connected and activation layers. In the standard workflow of Concrete, one provides the computation circuit to be compiled to an TFHE program as a Python function. Concrete supports a wide range of standard Python functions and a solid subset of the NumPy functionality\footnote{During our development, not all functions worked as expected, but choosing an equivalent alternative, modeling the functionality directly as a lookup table or subdividing complex broadcast operations into smaller sub-operations often solved compilation problems.}. We developed the components described in \cref{sec:training_scheme} as individual Python functions and composed them to the forward- and backward-passes of the individual layers.

\subsection{Tracing \& Compilation}
As part of the compilation process, Concrete traces the circuit to determine the value ranges of all intermediate results. \schemename{} pre-computes the compilation of the circuit during the input-independent offline phase using randomly generated input- and label-tensors.

Compiling large and complex circuits as required to train MLPs presents a challenge, as compilation times blow up because of excessive circuit-wide optimization. Compiling the whole for- and backward-pass into a single circuit is therefore not feasible. To circumvent the optimization overhead, we compiled the for- and backward-passes as well as weight updates of each layer and the error computation regarding the loss into separate circuits and composed them into a single training pass using Concrete's modules feature. \schemename{} benefits from compiler optimizations inside each of these components, while keeping overall compile times manageable. %
To train an MLP model for multiple epochs and data batches, the server evaluates the training pass repeatedly while always using the updated weights of the previous iteration (also see \cref{sec:threat_model}). We build a simple Keras-inspired interface to stack MLP-layers in Python by monkey-patching Concrete and enabling dynamic input-shapes for our layers.

\section{Evaluation}
Similar to previous work, we evaluate \schemename{} only regarding the online runtime performance and do not evaluate the communication costs of the inputs, as they are negligible in the context of the whole runtime overhead induced by FHE. To prevent stalling, in practice, one can start the training process after the first encrypted input batch arrives at the server, interlacing communication and computation.

In the following, we evaluate the performance of \schemename{} in terms of online runtime for eight real-world MLP training tasks and compare the prediction performance of the resulting models to standard PyTorch training using 32\,bit floating-point computations.

\subsection{Datasets, Models \& Hyperparameters}
\label{sec:datasets_models_hyperparameters}
We used the following datasets and MLP models for the evaluation. To validate the model performance, we used a 30\% test split for the thyroid cancer dataset and a 20\% test split for all other datasets. We leveraged uniform weight initialization, a signed output bitwidth $\Gamma=7$, $\operatorname{ReLU}_x$ with $x=14$ and a loss approximation level $\kappa=4$ for all models trained using \schemename. For the models trained with PyTorch, we used the default uniform weight initialization and classic ReLU activations. As we did not achieve the expected classification performance when training the PyTorch models using SGD \cite{robbins1951stochastic}, we use the Adam optimizer \cite{DBLP:journals/corr/KingmaB14} with a learning rate of $lr=0.001$. Similar to NITI \cite{DBLP:journals/tpds/WangRLS22}, we train all models without bias values. For simplicity, we used a batch size of eight samples for all models. To represent categorical features, we used a one-hot encoding. Additionally, we performed some standard preprocessing on all datasets, including removing the mean of some features, replacing zero values with the mean, and transforming skewed value distributions with a quantile transformer\footnote{To ensure reproducibility, the exact preprocessing procedure is part of our open-source artifact.}.
To simplify the circuit construction and reduce compile times, we drop the last data batch of each epoch if the number of its samples is smaller than the specified batch size. \cref{tab:dataset_summary} shows a summary of all used datasets and MLP configurations.

\begin{table}%
\centering
\caption{Summary of datasets: samples \#S, features \#F, classes \#C, input bitwidth $\beta$. Summary of MLPs: bitwidth of model parameters $\alpha$, epochs \#E, number of trainable parameters \#P and architectures.}
\label{tab:dataset_summary}
\setlength{\tabcolsep}{3.7pt}
\begin{tabular}{lcccccccc}
\toprule
Dataset & \#S & \#F & \#C & $\beta$ & $\alpha$ & \#E & \#P & MLP \\
\midrule
B. Cancer~\cite{breast_cancer} & 569 & 30 & 2 & 4 & 5 & 25 & 1080 & 28-8-2 \\
T. Cancer~\cite{borzooei2024machine} & 383 & 16 & 2 & 2 & 2 & 1 & 840 & 20-8 \\
Diabetes~\cite{smith1988using} & 768 & 8 & 2 & 4 & 5 & 25 & 144 & 8-8-2 \\
Wine~\cite{lichman2013uci} & 178 & 13 & 3 & 2 & 6 & 25 & 297 & 13-8-3 \\
V. Column~\cite{vertebral_column_212} & 310 & 6 & 3 & 5 & 7 & 100 & 164 & 10-8-3 \\
Parkinsons~\cite{parkinsons_174} & 197 & 22 & 2 & 2 & 6 & 50 & 912 & 32-8-2 \\
H. Disease~\cite{heart_disease_45} & 303 & 13 & 2 & 2 & 5 & 50 & 289 & 13-8-2 \\
H. Failure~\cite{heart_failure_clinical_records_519} & 299 & 12 & 2 & 2 & 6 & 50 & 256 & 12-8-2 \\

\bottomrule
\end{tabular}
\end{table}

\subsection{End-to-End Runtime}
\label{sec:end_to_end_runtime}
To evaluate the online runtime of \schemename{}, we follow the approach of previous work, e.g., like \cite{DBLP:conf/cvpr/NandakumarRPH19} and measure the runtime of our FHE-protected implementation of \schemename{} for a single data batch and subsequently project it to the whole runtime, meaning the number of batches required to achieve the best respective PyTorch-level test accuracy. To ensure the correctness of the FHE-protected computations, we compared the resulting activations, errors, weight gradients, and the final weights of all layers with our NumPy-based implementation.

As neither FHESGD \cite{DBLP:conf/cvpr/NandakumarRPH19} nor Glyph \cite{DBLP:conf/nips/LouFF020} provides an openly available implementation and both schemes only achieve 80\,bit of security, thus being insecure, we refrain from a comparison. Using FHE parameters to achieve 80\,bit security in the implementation of \schemename{} to ``enable'' a comparison is not useful as it would not be transferable to a secure setting\footnote{Concrete only supports secure execution with 128-bit security}. \Cref{tab:training_performance} reports the runtime results for all eight MLPs. As expected, the overhead induced by the FHE-protection with 128\;bit security is substantial, but we argue it is still manageable and, in many scenarios, practical. While being orders of magnitude slower than clear text training, \schemename{} achieves security guarantees compelling for the outsourcing of training in many use cases handling highly sensitive data, like in the medical domain. In these settings, \schemename{} can enable the non-interactive outsourcing of training for the first time. Also, compared to TEE-based outsourced training, \schemename{} introduces considerable overhead. But again, we argue \schemename{} achieves much stronger security guarantees by never decrypting on the server-side and keeping the decryption key only stored on the client side, not exposing it to potential server-side side-channel attacks. Thus, \schemename{} can enable non-interactive outsourcing of training in scenarios, where TEE-based security guarantees are not suitable.

\subsection{Layer-wise Runtime Distribution}
\label{sec:layer_wise_runtime_distribution}
\begin{figure}
	\centering
	\includegraphics[width=\linewidth]{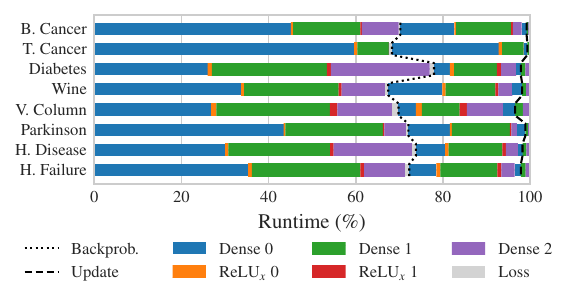}
	\caption{Runtime distribution over the individual circuits composed to form \schemename's end-to-end training pass for the eight MLP models of our evaluation. The vertical lines indicate at which point the backpropagation and the weight update phases begin.}
	\label{fig:mlp_runtime_dist}
\end{figure}

\Cref{fig:mlp_runtime_dist} shows the runtime distribution over the individual circuits composing \schemename's training pass for all eight evaluated MLP models. The first thing to notice is that the large majority of the runtime is spent on the forward- and backward passes of the dense layers, while the activation functions, loss computations, and weight updates only insignificantly contribute to the end-to-end runtime in comparison.
Secondly, it is noticeable that the forward-pass requires more than 60\% and often close to 80\% of the end-to-end runtime, although the backward-pass has to perform two matrix multiplications for the error and the weight gradient computations.
We attribute the efficiency of the backward-pass to its much smaller resolution requirements and \schemename's ability to adaptively leverage much smaller RNS representations and efficient RNS-based sign extraction gadget $\operatorname{Sign}_{(-1,0,1)}$.
Based on these results, future research should focus on further accelerating the linear layers to reduce the computational overhead of FHE-protected non-interactive outsourced training over its clear-text pendant.

\subsection{Classification Accuracy}
\label{sec:classification_accuracy}
To understand \schemename's ability to train accurate MLP models, we trained all eight MLPs using our fast clear-text NumPy-Implementation of \schemename{} and PyTorch. We report the best test accuracies achieved during the training in \cref{tab:training_performance}. \schemename{} achieves the same or better test accuracies for all datasets compared to the PyTorch-based 32-bit floating point training.

\begin{figure*}[t!]
	\centering
	\includegraphics[width=\linewidth]{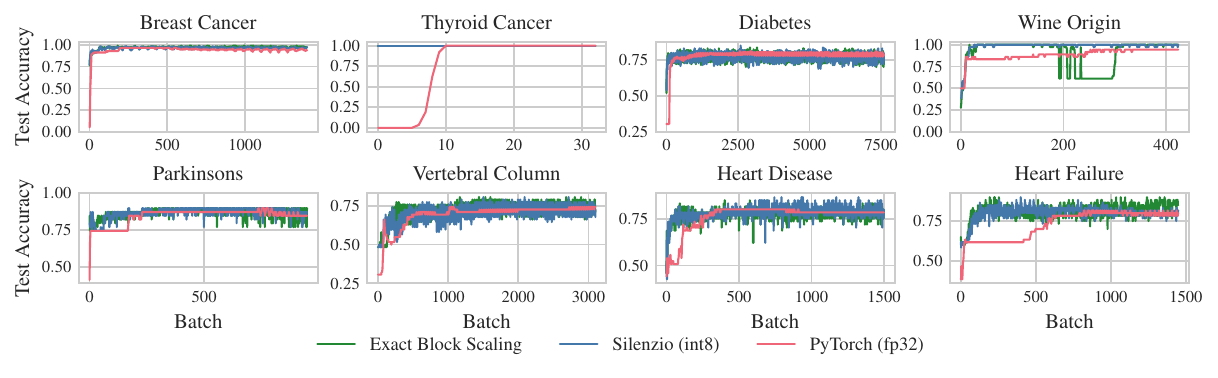}
	\caption{Test accuracy of MLPs trained using \schemename{} with $\text{Shift2MSBs}^{\pm}$ or exact block-scaling compared to PyTorch (fp32).}
	\label{fig:train_accs}
\end{figure*}

\Cref{fig:train_accs} compares the development of the test accuracy during the training with \schemename{} to the training using PyTorch. Interestingly, \schemename's training  converges faster than PyTorch in all cases.
The accuracy developments of all eight models clearly demonstrate the effective learning ability of our low-bitwidth integer based training approach. To recap, \schemename{} achieves the same or better classification accuracy compared to PyTorch in all evaluated classification tasks, being a secure and accurate drop-in replacement for unprotected training while creating a solid foundation for future research aiming to further improve the effectiveness of integer-based non-interactive outsourced training.

\begin{table}
\setlength{\tabcolsep}{1pt}
\centering
\caption{\schemename's runtime performance and training performance in comparison to standard PyTorch. We report the projected runtime required to achieve the best (left) and the PyTorch-level (right) test accuracy.}
\label{tab:training_performance}
\begin{tabular}{lcccc}
\toprule
Model & \makecell{Test Acc.\\PyTorch (fp32)} & \makecell{Test Acc.\\\schemename{}} & \makecell{Time\\(s/batch)} & \makecell{Projected\\time (h)} \\\midrule
B. Cancer & 96.5\% & \textbf{98.3\%} & 1027 & 19.4 / 17.7\\
T. Cancer & \textbf{100.0\%} & \textbf{100.0\%} & 451 & 0.1 / 0.1\\
Diabetes & 81.2\% & \textbf{85.1\%} & 295 & 212.6 / 31.7\\
Wine & 94.4\% & \textbf{100.0\%} & 419 & 3.1 / 2.0\\
V. Column & 74.2\% & \textbf{80.7\%} & 450 & 187.8/ 25.3\\
Parkinsons & \textbf{89.7\%} & \textbf{89.7\%} & 924 & 43.6 / 43.6 \\
H. Disease & 80.3\% & \textbf{86.9\%} & 387 & 80.1 / 8.9 \\
H. Failure & 81.7\% & \textbf{90.0\%}  & 361 & 40.1 / 11.7 \\
\bottomrule
\end{tabular}
\end{table}

\begin{figure*}[t!]
	\centering
    \includegraphics[width=\linewidth]{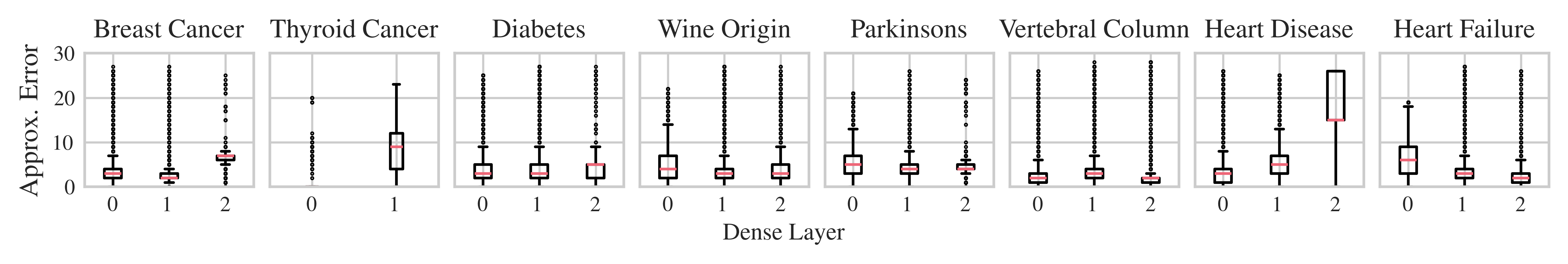}
	\caption{Approximation error distribution of \schemename's approximated block-scaling gadget during the training.}
	\label{fig:approx_error_boxplot}
\end{figure*}

\subsection{Approximation Impact}
\label{sec:approximation_impact}
We quantify the approximation error of $\text{Shift2MSBs}^{\pm}$ by tracking its absolute difference to an exact block-scaling implementation during training. \Cref{fig:approx_error_boxplot} shows the error distribution of the dense layer activations during the training.
While the error deviates significantly between layers and datasets, the mean of the error stays below ten for nearly all layers and MLPs except for the last layer of the heart disease MLP, which underlies, in general, large errors. On average the absolute error is 4.7, which, considering that we trained all models with a signed output bitwidth $\Gamma=7$, equates to a relative error of 3.7\%. The extreme error cases are substantial, with up to an absolute value of 28, or a relative error of 21.9\%.

Next, we trained all MLP models with the exact block-scaling implementation for comparison. As shown in \cref{fig:train_accs}, the training progress of \schemename{} with the approximated scaling gadget is, despite the partly large approximation errors, very similar compared to training using the exact block scaling implementation. To conclude, we found, that the approximation error of \schemename's new block scaling can be extreme on edge cases but is relatively small on average and does not affect the progress of the evaluated MLP training tasks. %

\section{Related Work}
\label{sec:related_work}
In their seminal work, \etal{Nandakumar} \cite{DBLP:conf/cvpr/NandakumarRPH19} proposed the first non-interactive outsourcing scheme for the training of MLPs based on the BGV FHE-scheme. The approach uses 8\,bit integer inputs and 16\,bit integer weights. They use sigmoid activations and optimize a quadratic loss using SGD during training. The evaluations showed the successful training of a model for the MNIST dataset, achieving 96\% accuracy. Measured by today's standards, the scheme is not secure by assuring only 80\,bit of security. Unfortunately, the scheme also lacks an openly available implementation.

Glyph \cite{DBLP:conf/nips/LouFF020} proposed by \etal{Lou} introduces a scheme-switching technique between TFHE and BGV to accelerate the non-interactive training of MLPs by performing non-linear activation functions like ReLU and softmax using TFHE and multiply accumulate operations protected through BGV. Further, Glyph demonstrates the fine-tuning of a CNN which brings two performance benefits. First, Glyph heavily reduces the number of trainable parameters; second, operations on the fixed parameters can be performed between ciphertexts and plaintexts, which is generally much faster. Note, that compared to \cite{DBLP:conf/provsec/JinRA20, DBLP:conf/icml/LeeLKSL23}, the client does not need to perform the feature extraction locally when using Glyph, making the scheme much more suitable for outsourcing. Glyph leverages a $L^2$ norm loss function implemented in BGV and quantized inputs, weights, and activations using SWALP training quantization \cite{DBLP:journals/corr/abs-1904-11943}. Glyph only achieves a security level of 80\,bit and does not provide an openly available implementation.

\etal{Montero} introduced a quantized NN training scheme based on TFHE, which we refer to as TFHE-T \cite{DBLP:conf/codaspy/MonteroFKBS24}. In TFHE-T, after each weight update, the server sends all weights to the client for re-encryption, making the scheme interactive. The authors suggest using bootstrapping operations instead to make the scheme non-interactive, but haven't demonstrated the adaption in an implementation and didn't evaluate it. TFHE-T was configured to achieve a security level of 128\,bit. The evaluation demonstrates the training of logistic regression models and MLPs with one hidden layer for the mortality (10 features) and breast-cancer datasets (30 features), achieving near-to plaintext training accuracy.
As the authors did not discuss their network settings, it seems they evaluated their interactive approach on a single machine, neglecting the communication overhead over the network, making comparisons to other schemes unfair.
To the best of our knowledge, the authors do not provide an open-source implementation.

PrivGD \cite{DBLP:conf/provsec/JinRA20} and Hetal \cite{DBLP:conf/icml/LeeLKSL23} are based on the CKKS FHE scheme and provide solutions for non-interactive fine-tuning of a single fully connected layer with an approximated Softmaxfunction on the output. Both schemes require the client to perform local feature extraction on the client-side only partially outsourcing the NN computations. \etal{Bourse} \cite{DBLP:conf/crypto/BourseMMP18} proposed a scheme for non-interactive inference on discretized MLPs. Furthermore, there exists a long line of research studying non-interactive CNN inference \cite{DBLP:conf/icml/Gilad-BachrachD16, DBLP:journals/corr/abs-1811-09953, DBLP:conf/icml/BrutzkusGE19, DBLP:conf/icml/LeeLLK0NC22, DBLP:journals/access/LeeKLCEDLLYKN22, DBLP:conf/ndss/FolkertsGT23, DBLP:conf/uss/AoB24, DBLP:conf/uss/Cheon00LJKL024, DBLP:conf/ccs/JuP0KKCA24, DBLP:journals/ijns/RovidaL24, DBLP:journals/popets/KuLHCHTC25}. \etal{Zhang} \cite{DBLP:conf/nips/ZhangZSCL24} presented HEPrune for non-interactive (still requiring light-weight meta-data online-communication) FHE protected data pruning. DataSeal \cite{DBLP:journals/corr/abs-2410-15215} adds integrity to FHE schemes, making them secure against malicious attackers and demonstrated the protection mechanism on NN inference.

Following the latest advancements in natural language processing, \etal{Zhang} \cite{DBLP:journals/iacr/ZhangLYWCHRY24} proposed non-interactive transformer inference and \etal{Panzade} \cite{DBLP:journals/corr/abs-2402-09059} introduced BlindTuner for non-interactive fine-tuning (using client-side local feature extraction) of a transformer model. Furthermore, there are solutions for k-NN classification \cite{DBLP:journals/popets/ZuberS21}, tree inference \cite{DBLP:conf/ccs/Cong00P22, DBLP:conf/ccs/MahdaviNLK23}, 
SVM and K-Means classification \cite{DBLP:journals/iacr/BianZSZMLLWSGHJL24}
for the non-interactive outsourcing setting.

\section{Future Work}
Although \schemename{} demonstrates a big step towards secure non-interactive outsourced MLP training, several directions remain open to further improve performance, generality, and practical deployment. To accelerate the FHE execution, Concrete supports compiling for GPU and FPGA targets. Unfortunately, at the time of writing, the compiler backends for CPU and GPU are seemingly unequally mature and our code only compiled successfully for CPU execution. For future research, it would be interesting to also consider hardware accelerators like GPUs and \schemename's implementation gives a good starting point, presumably already supporting GPU execution in a future release of Concrete's GPU compiler.

Next to the MLPs supported by \schemename, future research should tackle the challenge of supporting more complex architectures, like large CNNs and recurrent models with support for more of the layers used in standard models. \schemename's  $\text{Shift2MSBs}^\pm$ block-scaling gadget can also compute the approximated amount of shift performed on the input tensor, this might be helpful to construct fixed-point approximations with higher resolution, similar to NITI \cite{DBLP:journals/tpds/WangRLS22}, required to successfully train larger and more complex model architectures.

Finally, more research is required to further strengthen the security guarantees provided by non-interactive training schemes. To the best of our knowledge, there exists no training scheme providing security against a malicious attacker. In the semi-non-interactive outsourcing setting, \cite{DBLP:conf/uss/KotiPPS21, DBLP:journals/popets/WaghTBKMR21, DBLP:journals/popets/AttrapadungHIKM22, DBLP:conf/ndss/KotiPRS22} achieve security against a malicious attacker, but require non-collusion assumptions. Dataseal \cite{DBLP:journals/corr/abs-2410-15215} provides verifiability for FHE protected computations and demonstrated non-interactive outsourced inference, which could serve as the basis for the development of a training scheme.

\section{Conclusion}

Based on careful co-design of an approximated MLP training scheme and low-bitwidth integer components efficiently realizable in FHE, we presented \schemename{}. To the best of our knowledge, \schemename{} is the first fully non-interactive framework for outsourcing the training of MLPs, which demonstrates the practical realizability of a 128\,bit security guarantee. At its core are three novel building blocks — a low-bitwidth matrix-multiplication gadget, the $\text{Shift2MSBs}^\pm$ block-scaling mechanism, and an integer-only cross-entropy gradient computation — which together enable a true “fire-and-forget” training paradigm without interaction or non-collusion assumptions. While the non-interactive outsourcing of inference workloads is a well-researched problem, we hope \schemename{} will serve as an impulse for further advancements regarding the less researched and much harder training task.

Our end-to-end implementation in Zama’s Concrete library demonstrates that \schemename{} achieves test accuracies on par with standard PyTorch, leveraging 32\,bit floating-point computations, while incurring manageable FHE runtime overheads.
By removing both client–server interactions and collusion assumptions, \schemename{} offers a drop-in replacement for conventional cloud training services, unlocking non-interactive privacy-preserving outsourced training for sensitive domains such as healthcare and finance. Finally, we will open source \schemename's implementation, providing the first open available solution for non-interactive outsourced MLP training.

\section*{Acknowledgment}
Generative AI was utilized during programming, editing, and grammar enhancement of this work.
This work has been supported by the BMBF through the project AnoMed.

\appendices
\section{RNS-to-MRNS Conversion Example}
\cref{tab:rns_to_mrs_example} shows an example of the conversion process from a number given in RNS representation to the associated MRNS.

\begin{table}[h!]
    \centering
    \caption{Example of the RNS to MRNS conversion process.}
    \label{tab:rns_to_mrs_example}
    \begin{tabular}{l|cccc}
    Moduli $m_1, m_2, m_3$, Radices $r_1, r_2, r_3$             & 5 & 7 & 8\\
    RNS of $x=99$                                               & 4 & 1 & 3\\
    \midrule
    $(x_1, x_2-4, x_3-4)$                                       & 4 & 4 & 7\\
    $(x_1, x_2 \cdot 5^{-1}\bmod 7, x_3 \cdot 5^{-1}\bmod 8)$   & 4 & 5 & 3\\
    \midrule
    $(x_1, x_2, x_3-5)$                                         & 4 & 5 & 6\\
    $(x_1, x_2, x_3\cdot 7^{-1}\bmod 8)$                        & 4 & 5 & 2\\
    \midrule
    MRNS of $x=4+5\cdot5+2\cdot7\cdot5=99$                      & 4 & 5 & 2\\
    \end{tabular}
\end{table}

\section{Low-bitwidth high-resolution MatMul}
\label{sec:low_bw_high_res_matmul}
\begin{figure}
	\centering
    \includegraphics[width=\linewidth]{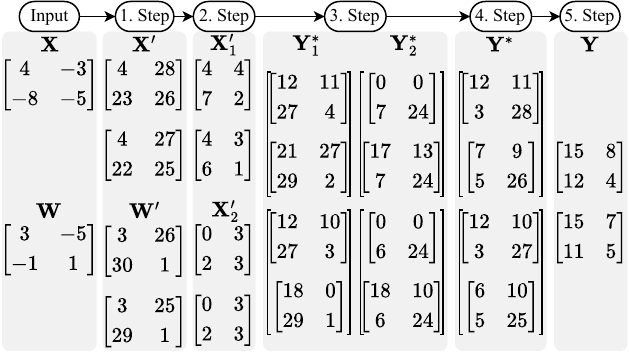}
	\caption{Minimal step-by-step example of \cref{lst:matmul_high_res} for 5-bit RNS-base $[31, 30]$.}
	\label{fig:matmul_example}
\end{figure}

Similar to $\operatorname{Matmul}^{\text{RNS}}$, $\operatorname{MatmulHighRes}^{\text{RNS}}$ (see \cref{lst:matmul_high_res}) gets up to 8\,bit matrices as inputs, but the used RNS bases in $\operatorname{MatmulHighRes}^{\text{RNS}}$ consist of larger 5\,bit moduli (see \cref{tab:rns_bases_5bit}). The first step is identical in both algorithms. To account for the larger RNS moduli, the second step draws inspiration from the Karatsuba algorithm to keep the intermediate results of the following computations also $\leq$ 8\,bit and extracts individually the most significant two and least significant three residue-bits of the first input matrix in RNS. In step three, the partial products $\mathbf{Y}_1^*$ and $\mathbf{Y}_2^*$ are computed, where $\mathbf{Y}_1^*$ is immediately reduced modulo the RNS base. To respect the bit extractions of step 2 and perform the modular reduction of $\mathbf{Y}_2^*$ without exceeding 8\,bit, we perform the evaluation using a custom lookup table. The last two steps are identical to \cref{lst:matmul}, except, that we respect the larger intermediate results and only sum over maximal 7 summands in step 5. \cref{fig:matmul_example} shows a step-by-step example of $\operatorname{MatmulHighRes}^{\text{RNS}}$.

\begin{algorithm}
\caption{$\operatorname{MatmulHighRes}^{\text{RNS}}$}
\label{lst:matmul_high_res}
\begin{algorithmic}
    \Require $\mathbf{X} \in \mathcal{I}_8^{a \times b}$, $\mathbf{W} \in \mathcal{I}_8^{b \times c}$, $\mathbf{m} \in \mathcal{U}_5^k$ \MyComment{Inputs, RNS base}
    \Ensure $\mathbf{Y} \in \mathcal{J}_5^{k \times a \times b}$ \MyComment{Output in RNS representation}

    \LineComment{1. Step: Convert inputs to RNS representation}
    \State $\mathbf{m}' \gets \operatorname{expandDims}(m, [1, 2])$ \MyComment{Shape: $k \times 1 \times 1$}
    \State $\mathbf{m}'' \gets \operatorname{expandDims}(m, [1, 2, 3])$ \MyComment{$k \times 1 \times 1 \times 1$}
    \State $\mathbf{X}' \gets \mathbf{X} \bmod \mathbf{m}'$ \MyComment{$k \times a \times b$}
    \State $\mathbf{W}' \gets \mathbf{W} \bmod \mathbf{m}'$ \MyComment{$k \times b \times c$}

    \LineComment{2. Step: Extract bitchunks of input $X'$}
    \State $\mathbf{X}_2' \gets \operatorname{extractBits}(\mathbf{X}',[3, 4])$ \MyComment{$k \times a \times b$}
    \State $\mathbf{X}_1' \gets \operatorname{extractBits}(\mathbf{X}',[0, 2])$

    \LineComment{3. Step: Compute partial products}
    \State $\mathbf{X}_1' \gets \operatorname{expandDims}(\mathbf{X}', 3)$ \MyComment{$k \times a \times b \times 1$}
    \State $\mathbf{X}_2' \gets \operatorname{expandDims}(\mathbf{X}', 3)$
    \State $\mathbf{W}' \gets \operatorname{expandDims}(\mathbf{W}', 1)$ \MyComment{$k \times 1 \times b \times c$}
    \State $\mathbf{Y}_1^* \gets \mathbf{X}_1' \mathbf{W}' \bmod m''$ \MyComment{$k \times a \times b \times c$}
    \State $\mathbf{Y}_2^* \gets \mathbf{X}_2' \mathbf{W}'$
    \State $\mathbf{Y}_2^* \gets \operatorname{lookupComp}(\mathbf{Y}_2^*\cdot 8 \bmod m'')$

    \LineComment{4. Step: Add partial products}
    \State $\mathbf{Y}^* \gets \mathbf{Y}_1^* + \mathbf{Y}_2^* \bmod m''$ 

    \LineComment{5. Step: Block-wise summation}
    \State $n \gets \min(7,b)$
    \State $\mathbf{Y} \gets \operatorname{sum}(\mathbf{Y}^*[:, :, :n, :], axis=2) \bmod \mathbf{m}'$

    \For{$i \gets n,\; i \le b,\; i \gets i+n$}
        \State $j \gets \min(i+n, b)$
        \State $\mathbf{Y}' \gets \operatorname{sum}(\mathbf{Y}[:,:,i:j,:], axis=2)$
        \State $\mathbf{Y} \gets \mathbf{Y} + \mathbf{Y}' \bmod \mathbf{m}'$ \MyComment{$k \times a \times c$}
    \EndFor
    
\end{algorithmic}
\end{algorithm}

\begin{table}
\centering
\caption{5\,bit RNS bases.}
\label{tab:rns_bases_5bit}
\begin{tabular}{llc}
\toprule
$k$ & RNS Base & Max. Bitwidth\\\midrule
2 & 31, 30 & 9.86\\
3 & 29, 31, 30 & 14.72\\
4 & 27, 29, 31, 28 & 19.37\\
5 & 25, 27, 29, 31, 28 & 24.02\\
6 & 23, 25, 27, 29, 31, 28 & 28.54\\
\bottomrule
\end{tabular}
\end{table}

\bibliographystyle{IEEEtran}
\bibliography{bib.bib}

\end{document}